\documentclass{ws-procs9x6}
\begin{document}

\title{Microscopic Origin of the Shear Relaxation Time in Causal Dissipative
Fluid Dynamics}
 
\author{Gabriel S.\ Denicol}  

\address{Institut f\"ur Theoretische Physik, Goethe University, 60438
Frankfurt am Main, Germany}

\author{Harri Niemi}

\address{Frankfurt Institute for Advanced Studies (FIAS),
60438 Frankfurt am Main, Germany}

\author{Jorge Noronha}  

\address{Instituto de F\'{i}sica, Universidade de S\~{a}o Paulo, C.P. 66318,
05315-970 S\~{a}o Paulo, SP, Brazil}
\author{Dirk H.\ Rischke } 

\address{Institut f\"ur Theoretische Physik, Goethe University, and
\\
Frankfurt Institute for Advanced Studies (FIAS), 60438 Frankfurt am Main, Germany }

\begin{abstract}
In this paper we show how to compute the shear relaxation 
time from an underlying microscopic theory.
We prove that the shear relaxation time in Israel-Stewart-type theories 
is given by the inverse of the pole of the 
corresponding retarded Green's function, which is 
nearest to the origin in the complex energy plane. Consequently, the
relaxation time in such theories is a microscopic, and not a 
macroscopic, i.e., fluid-dynamical time scale. 
\end{abstract}


\bodymatter

\section{Introduction}

Relativistic fluid dynamics has been applied with success to describe the
dynamics of hot and dense matter created in relativistic heavy-ion
collisions at RHIC, and recently, at LHC \cite{Huovinen:2006jp}. Despite
this success, the derivation of relativistic fluid dynamics from an
underlying microscopic
theory is still an open problem and is under intense theoretical
investigation. This is mainly due to nontrivial consequences imposed by
causality and stability in relativistic fluids \cite{hiscock,us}.

In this paper we discuss a general method to compute all linear
transport coefficients (such as the shear viscosity and its corresponding
relaxation time) associated with the dissipative behavior of
fluids. In particular, we prove that this prescription gives a value for the
shear relaxation time that, under the 14-moment approximation, coincides
with the one derived from kinetic theory. This shows that transient dynamics
is determined by the slowest microscopic and not by the fastest
fluid-dynamical time scale. Our approach \cite{DNNR}
depends only on generic analytical properties of retarded Green's
functions.

\section{Definitions and Power-Counting Scheme}

Let us consider the following linear relation between a dissipative current $%
J\left( X\right) $ and its thermodynamical force $F\left( X\right) $ in
Fourier space, 
\begin{equation}
\tilde{J}\left( Q\right) =\tilde{G}_{R}\left( Q\right) \tilde{F}\left(
Q\right) ,  \label{mainEq}
\end{equation}%
where we define the Fourier transformation in the following way, 
\begin{eqnarray}
\tilde{A}\left( Q\right)  &=&\int d^{4}X\exp \left( iQ\cdot X\right) A\left(
X\right) \;, \\
A\left( X\right)  &=&\int \frac{d^{4}Q}{\left( 2\pi \right) ^{4}}\exp \left(
-iQ\cdot X\right) \tilde{A}\left( Q\right) \;.
\end{eqnarray}%
Here, we adopt the following notation for the four-momentum and space-time coordinate, $Q = (\omega,\mathbf{q})$ and $X = (t,\mathbf{x})$, respectively.
In this article, we discuss particle diffusion, in which $J$ is the
diffusion current of a density $n$ and $F\sim \partial n$, and shear flow,
in which $J$ is the shear stress tensor $\pi ^{\mu \nu }$ and $F$ the shear
tensor $\sigma ^{\mu \nu } \equiv \nabla^{\langle \mu} u^{\nu \rangle}$,
with the fluid four-velocity $u^\mu$, $\nabla^\mu \equiv \Delta^{\mu \nu}
\partial_\nu$, $\Delta^{\mu \nu} = g^{\mu \nu} - u^\mu u^\nu$, and
$A^{\langle \mu \nu \rangle} \equiv \frac{1}{2}(\Delta^{\mu \alpha}
\Delta^{\nu \beta} + \Delta^{\mu \beta} \Delta^{\nu \alpha} - \frac{2}{3}\,
\Delta^{\mu \nu} \Delta^{\alpha \beta}) A_{\alpha \beta}$.

Equation (\ref{mainEq}) 
contains all the information about the linearized underlying
microscopic theory in the form of the retarded Green's function
$\tilde{G}_R(Q)$. When the
system exhibits a clear separation between the typical microscopic and
macroscopic scales, $\lambda $ and $\ell $, respectively, it is generally
assumed that the dynamics of the system can be described in terms of a
finite number of macroscopic variables. The macroscopic scale is associated
with the variations of the thermodynamic force $F\sim \ell ^{-1}$ and its
derivatives $\partial^{n}F\sim \ell ^{-n-1}$. The microscopic scale $\lambda$ 
is
contained in the poles and derivatives of $\tilde{G}_{R}(Q)$ and would be, for
example, the mean-free path in dilute gases. The challenge is to determine
the equation of motion satisfied by the current $J$ in terms of the
microscopic information contained in the 
retarded Green's function $\tilde{G}_{R}(Q)$ 
when such a separation of scales exists.

\section{Gradient Expansion}

The separation between microscopic and macroscopic scales is normally 
used to determine the macroscopic
dynamics of the dissipative current $J$ in terms of an expansion in powers
of Knudsen number, \textrm{K}$\mathrm{n}\equiv \lambda /\ell \ll 1$. The
simplest realization of this idea is known as the gradient expansion \cite%
{Chapman}. In the gradient expansion the dissipative currents are expressed
solely in terms of gradients of fluid-dynamical variables. In practice, the
gradient expansion is implemented by expanding the retarded 
Green's function, $\tilde{G}_{R}(Q)$, in a Taylor series 
around the origin, $Q=0$, 
\begin{equation}
\tilde{G}_{R}( Q) \sim \tilde{G}_{R}( 0,\mathbf{0}) +\partial_{\omega }
\tilde{G}_{R}( 0,\mathbf{0}) \, \omega +\ldots \,,
\end{equation}%
where the terms resulting from the expansion in $\mathbf{q}$ are not 
explicitly denoted. 
This expansion leads to the following equation of motion for $J$%
\begin{equation}
J\left( X\right) =CF\left( X\right) +C_{1}\partial _{t}F\left( X\right)
+\ldots \,,  \label{gradient1}
\end{equation}%
where $C=\tilde{G}_{R}\left( 0,\mathbf{0}\right) $, 
$C_{1}=\partial _{\omega }\left.
\tilde{G}_{R}\left( Q\right) \right\vert _{\omega ,\mathbf{q}=0}$, $\ldots $. 
We
remark that the gradient expansion usually does not involve time derivatives
of the thermodynamic force. This is not a problem since one can in principle
use the conservation laws to replace time derivatives by spatial gradients.

Apart from overall factors which ensure the correct dimensionality,
$C \sim \lambda$ and $C_{1} \sim \lambda^2$, respectively.
Thus, the term $CF$ is of first
order in \textrm{K}$\mathrm{n}$ while $C_{1}\partial _{t}F\left( t,x\right) $
is of second order in the Knudsen number, \textrm{K}$\mathrm{n}^{2}$. If 
\textrm{K}$\mathrm{n}\ll 1$, higher-order gradients can in principle be
omitted and one obtains a closed macroscopic theory for $J$. Navier-Stokes
theory \cite{landau} corresponds to truncating the series at first order
while keeping higher-order corrections leads to the Burnett equations \cite%
{Burnett}.

This method, however, leads to unstable fluid-dynamical theories. In the
non-relativistic case, the second-order truncation, i.e., the Burnett
equation, is unstable \cite{Bobylev}. In the relativistic case,
Navier-Stokes theory itself is unstable \cite{hiscock,us}. Recently, it has
been shown that this instability in relativistic theories is intimately
related to the violation of causality \cite{us}. Such problems prohibit the
application of these theories to describe any realistic system.

\section{Computing the Relaxation Time from the Poles of $\tilde{G}_{R}$}

In the derivation of Eq.\ (\ref{gradient1}), two main assumptions were
employed: the separation between the microscopic and macroscopic scales and
the gradient expansion itself. The first assumption is essential for the
coarse-graining procedure that defines fluid dynamics and cannot be
disregarded. On the other hand, the second assumption is solely based on the
idea that the Taylor expansion is able to describe the retarded Green's
function in the whole macroscopic/mesoscopic frequency domain. This need
not be true, since it is known that Green's functions are not
necessarily analytic in the whole complex plane. In fact, these functions
can have singularities in the complex $\omega $ plane, which necessarily
restricts the domain of validity of the Taylor expansion.

In order to illustrate the physical meaning of such 
singularities, it is convenient
to consider the generalized diffusion equation for $J$, also known as the
Maxwell-Cattaneo equation \cite{Cattaneo}. In this case, as in the
Israel-Stewart theories of fluid dynamics \cite{IS}, the dissipative current
satisfies a relaxation equation. In the linear domain, this can be written
as,%
\begin{equation}
\tau _{R}\frac{\partial }{\partial t}J+J=DF.  \label{Cattaneo}
\end{equation}%
In these theories there are two transport coefficients: the relaxation time $%
\tau _{R}$ and the diffusion coefficient $D$. In Fourier space, Eq. (\ref%
{Cattaneo}) can be written in the form $\tilde{J}( \omega ) =%
\tilde{G}_{R}( \omega ) \tilde{F}( \omega ) $, where%
\begin{equation}
\tilde{G}_{R}( \omega ) =\frac{D}{\tau_{R}}\,\frac{i}{\omega +
i/\tau_{R}}.
\end{equation}
It is important to notice the existence of a pole of $\tilde{G}_{R}
(\omega)$ on the
imaginary axis. The existence of this pole is directly
related to the existence of a non-zero relaxation time and, consequently, to
the causal structure of the theory \cite{us}. It is also clear from Eq.\ (\ref%
{Cattaneo}) that Fick's law and, consequently, the diffusion equation are
recovered in the limit of vanishing relaxation time, $\tau_{R}\rightarrow 0$. 
Mathematically, this limit pushes the pole to infinity, rendering 
$\tilde{G}_{R} (\omega)$ an analytic function in the 
whole complex $\omega $ plane. Only in
this case a Taylor expansion near the origin has an infinite radius of
convergence and is able to provide a good description of 
$\tilde{G}_{R}(\omega)$.

However, in transient theories of fluid dynamics, where the relaxation time
is necessarily nonzero, can a Taylor expansion still be used? Strictly
speaking, the range of the Taylor expansion is limited by the existence of
the pole. As was previously shown, the pole is directly related to the
relaxation time and, consequently, the Taylor expansion is unable to probe
time scales of the order of the relaxation time.

\section{The Role of the Analytical Structure of $\tilde{G}_{R}\left( 
\protect\omega \right) $}

Next, we consider a general analytical
structure for $\tilde{G}_{R}( Q)$. We assume that $\tilde{G}_{R}(Q)$ 
has $N$ poles in the complex $\omega $ plane, $\omega_{1}( 
\mathbf{q})$, \ldots , $\omega_{N}( \mathbf{q}) $, 
while the structure in the $\mathbf{q}$ plane is analytic. Note
that, in principle, the number of singularities does not need to be finite.
Also, in many cases $\tilde{G}_{R}( Q) $ contains branch cuts,
especially in the UV region, $\omega /\omega_{1}\gg 1$, which is sensitive
to temperature-independent properties of the system. We only consider the
region in the complex $\omega $ plane without branch cuts. Then, 
$\tilde{G}_{R}(Q)$ can be expressed in the following way%
\begin{equation}
\tilde{G}_{R}( Q) =\sum_{i=1}^{N}\frac{f_{i}(Q) }{%
\omega -\omega_{i}( \mathbf{q}) }=\frac{\Xi (Q) }{%
\left[ \omega -\omega_{1}( \mathbf{q}) \right] \cdots \left[
\omega -\omega_{N}( \mathbf{q}) \right] }\;,  \label{blabla}
\end{equation}%
where $f_{i}(Q) $ and $\Xi (Q) $ are analytic
functions in the complex $\omega $ plane. We write the right-hand side of
Eq.\ (\ref{blabla}) in a polynomial form,
\begin{equation}
1+\Phi _{1}( \mathbf{q}) \left( -i\omega \right) +\ldots +\Phi
_{N}( \mathbf{q}) \left( -i\omega \right)^{N}=\left( -1\right)^{N}
\frac{\left[ \omega -\omega_{1}(\mathbf{q}) \right] \cdots 
\left[ \omega -\omega_{N}( \mathbf{q}) \right] }{\omega_{1}
( \mathbf{q}) \cdots \omega_{N}(\mathbf{q}) }\;,
\label{poly}
\end{equation}
where the $\omega_{i}( \mathbf{q}) $ are the zeros of the
polynomial on the left-hand side and we introduced the functions $\Phi
_{m}( \mathbf{q}) $ 
\begin{equation}
\Phi _{m}( \mathbf{q}) =\left( -i\right)^{m}\sum_{1\leq
i_{1}...<i_{m}\leq N}\frac{1}{\omega _{i_{1}}( \mathbf{q}) \cdots
\omega_{i_{m}}( \mathbf{q}) }\;.
\end{equation}
Combining Eqs.\ (\ref{blabla}) and (\ref{poly}), we obtain
\begin{equation}
\left[ 1+\Phi _{1}( \mathbf{q}) \left( -i\omega \right) +\ldots
+\Phi _{N}( \mathbf{q}) \left( -i\omega \right) ^{N}\right] 
\tilde{G}_{R}( Q) =\frac{\left( -1\right)^{N}\Xi ( Q) 
}{\omega _{1}( \mathbf{q}) \cdots \omega_{N}( \mathbf{q}) }\;.  \label{bla}
\end{equation}
After taking the inverse Fourier transform and expanding the functions $\Phi
_{m}( \mathbf{q}) $ in a Taylor series around $\mathbf{q}=0$, we
obtain a differential equation satisfied by $G_{R}(X-Y)$. It should be
clear that this will be a linear differential equation of order $N$ in time.
The equation of motion for $J$ can be obtained by multiplying Eq.\ (\ref{bla}%
) by $\tilde{J}$\ and taking the inverse Fourier transform. See Ref. \cite%
{DNNR} for details. The result is%
\begin{equation}
\chi _{N}\partial _{t}^{N}J+\ldots +\chi _{1}\partial _{t}J+J=D_{0}F+\ldots
+D_{N}\partial _{t}^{N}F+\mathcal{O}\left( \partial _{t}^{N+1}F,\partial _{%
\mathbf{x}}\right) .  \label{Sugoy}
\end{equation}%
Here, we omitted all terms involving spatial derivatives.
We introduced the transport coefficients $\chi _{m}$ and $D_{m}$,  
\begin{eqnarray}
\chi _{m} &=&\Phi _{m}( \mathbf{0}) =\left( -i\right)
^{m}\sum_{1\leq i_{1}...<i_{m}\leq N}\frac{1}{\omega _{i_{1}}( \mathbf{0})
 \cdots \omega _{i_{m}}( \mathbf{0}) }\;,
\label{very nice} \\
D_{m} &=&i^{m}\frac{\left( -1\right) ^{N}}{m!}\frac{\partial _{\omega
}^{m}\left. \Xi ( \omega ,\mathbf{0}) \right\vert _{\omega =0}}{%
\omega_{1}( \mathbf{0}) \cdots \omega_{N}( \mathbf{0}) }\;.
\end{eqnarray}

If there is a clear separation of scales, i.e., $%
\mathrm{Kn}\ll 1$, it becomes possible to simplify the right-hand side of 
Eq.\ (\ref{Sugoy}). In this case, $D_{0}F\gg D_{1}\partial_{t}F\gg
D_{2}\partial _{t}^{2}F$, and higher derivatives of $F$ can be 
neglected. However, in order to obtain a relaxation-type
equation of motion for $J$, additional approximations are required. For this
purpose, one must impose the limiting procedure, $\chi_{i}\rightarrow 0$, $%
i\geq 2$. In such a limiting procedure, all the poles of the retarded Green's
function except the pole nearest to the origin (in the following referred
to as the \textquotedblleft first pole\textquotedblright ) are pushed to
infinity \cite{DNNR}. Assuming that this can be done, we obtain the
following equation of motion for $J$%
\begin{equation}
\tau_{R}\partial_{t}J+J=D_{0}F+D_{1}\partial_{t}F+D_{2}\partial _{t}^{2}F+%
\mathcal{O}\left( D_{3}\partial _{t}^{3}F,\partial _{\mathbf{x}}\right) \;,
\label{38}
\end{equation}%
where one can show that 
\begin{eqnarray}
\tau_{R} &=&\frac{1}{i\omega _{1}( \mathbf{0}) },  \nonumber \\
D_{0} &=& \tilde{G}_{R}( 0 ,\mathbf{0}) \;,  \nonumber \\
D_{1} &=&i\partial_{\omega }\left. \tilde{G}_{R}( \omega ,\mathbf{0}) 
\right\vert _{\omega =0}+D_{0}\tau _{R}\;.  \label{very nice5}
\end{eqnarray}
Note that the time reversal symmetry of the retarded Green's function
requires the first pole to lie on the imaginary axis, such that $\tau_R$
is real, as it should be.
It is also important to remark that Eq.\ (\ref{38}) describes the long- 
(but not asymptotically long-) time 
evolution of the system. This is the physical
origin of the limiting procedure previously implemented and it is the only
limit in which a relaxation equation can be obtained and, consequently, a
relaxation time can be defined. If one wants to describe even shorter time
scales, it becomes necessary to include more terms $\sim \chi _{i}$ in the
discussion. Then, the equations of motion will become more complicated.
Naturally, the gradient expansion is recovered when all poles are pushed to
infinity and, in this sense, the gradient expansion can be interpreted as an
asymptotic solution of the more general equation of motion \cite{DNNR}.

Note that this limiting procedure is only allowed when the first pole lies
on the imaginary axis. On the other hand, if the first pole and, for reasons
of symmetry, its counterpart on the other side of the imaginary axis, have
nonzero real parts one cannot disregard one of the poles while keeping the
other. Then, the long-time evolution of the dissipative current will be
characterized by a second-order differential equation and will display
oscillatory motion. Therefore, in this case the fluid-dynamical motion
cannot be reduced to a simple relaxation equation, even in the
small-frequency domain.

\section{Application: The Linearized Boltzmann Equation}

The discussion presented above is general and can, in principle, be applied
to any type of theory. In this section, we apply this formalism to kinetic
theory. In particular, we calculate the shear viscosity and relaxation time
coefficients for a weakly coupled classical gas of hard spheres via the
Boltzmann equation \cite{DNNR}.

We perform all our calculations in the local rest frame of the fluid
element, where $u^{\mu }=(1,0,0,0)$. Since we restrict our discussion to
shear viscosity, we consider the simplified scenario in which the fluid does
not accelerate, $u^\mu \partial_\mu u^{\nu }\equiv 0$, 
and does not expand, $\partial_\mu u^\mu \equiv 0$, 
and where temperature, $T=\beta _{0}^{-1}$, and chemical
potential, $\mu =\alpha _{0}/\beta _{0}$, are constant. With all these
simplifications, the linearized Boltzmann equation can be written as \cite%
{DNNR}%
\begin{equation}
\partial _{t}\delta f_{\mathbf{k}}+\mathbf{v}\cdot \nabla f_{\mathbf{k}}-%
\hat{C}\delta f_{\mathbf{k}}=\mathcal{S}\left( X,K\right) ,
\label{linearboltzmann}
\end{equation}%
where $\delta f_{\mathbf{k}}=f_{\mathbf{k}}-f_{0\mathbf{k}}$, with $f_{0%
\mathbf{k}}=\exp \left( \alpha _{0}-\beta _{0}E_{\mathbf{k}}\right) $, $%
\mathbf{v}=\mathbf{k/}E_{\mathbf{k}}$, and $\mathcal{S}=\beta _{0}f_{0%
\mathbf{k}}E_{\mathbf{k}}^{-1}k^{\left\langle \mu \right. }k^{\left. \nu
\right\rangle }\sigma _{\mu \nu }$. Here, we
introduced the collision operator, $\hat{C}$, defined as%
\begin{eqnarray}
\hat{C}\delta f_{\mathbf{k}}&=&\frac{1}{\nu E_{\mathbf{k}}}\int dK^{\prime
}dPdP^{\prime }W_{\mathbf{kk}\prime \rightarrow \mathbf{pp}\prime }f_{0%
\mathbf{k}}f_{0\mathbf{k}^{\prime }} \; \notag \\ 
&& \times \left( \frac{1}{f_{0\mathbf{p}^{\prime }}}
\delta f_{\mathbf{p}^{\prime }}+%
\frac{1}{f_{0\mathbf{p}}}\delta f_{\mathbf{p}}-\frac{1}{f_{0\mathbf{k}%
^{\prime }}}\delta f_{\mathbf{k}^{\prime }}-\frac{1}{f_{0\mathbf{k}}}\delta
f_{\mathbf{k}}\right) ,  \label{collisionoperatordefinition}
\end{eqnarray}%
where $W_{\mathbf{kk}\prime \rightarrow \mathbf{pp}\prime }$ is the
transition rate, $\nu $ is the symmetry factor ($\nu =2$ for identical
particles) and $dK=d^{3}\mathbf{k}\,E_{\mathbf{k}}^{-1}/(2\pi )^{3}$.

Using Eqs.\ (\ref{linearboltzmann}) and (\ref{collisionoperatordefinition}),
it is possible to express the Fourier transform of the shear stress tensor, $%
\tilde{\pi}^{\mu \nu }( Q) $, in terms of the Fourier transform
of the shear tensor, $\tilde{\sigma}_{\alpha \beta }( Q) $, in
the form $\tilde{\pi}^{\mu \nu }=G_{R}^{\mu \nu \alpha \beta }\tilde{\sigma}%
_{\alpha \beta }$ \cite{DNNR}, where%
\begin{equation}
\tilde{G}_{R}^{\mu \nu \alpha \beta }\left( \omega ,\mathbf{q}\right) =\int
dK\,k^{\left\langle \mu \right. }k^{\left. \nu \right\rangle }\,
\frac{1}{i\omega-i\mathbf{v}\cdot \mathbf{q}-\hat{C}}\,
\beta _{0}\,f_{0\mathbf{k}}E_{\mathbf{k}%
}^{-1}k^{\left\langle \alpha \right. }k^{\left. \beta \right\rangle \;}.
\label{Exp}
\end{equation}%
In order to compute this Green's function, we define 
a function $B^{\alpha \beta }(Q,K) $ which satisfies%
\begin{equation}
\left( i\omega-i\mathbf{v}\cdot \mathbf{q} -\hat{C}\right) \,
B^{\alpha \beta }( Q,K) =\mathcal{S}%
( Q,K) =\beta _{0}E_{\mathbf{k}}^{-1}k^{\left\langle \alpha
\right. }k^{\left. \beta \right\rangle }f_{0\mathbf{k}}.  \label{equation}
\end{equation}%
As shown in the previous section, in order to extract the shear viscosity and
relaxation time coefficients, it is sufficient to consider the case $\mathbf{%
q}=0$. Then, the dependence of $B^{\alpha \beta }$ on $K$ can be expressed
via the following expansion,%
\begin{equation}
B^{\alpha \beta }( \omega ,K) =f_{0\mathbf{k}}\,k^{\left\langle
\alpha \right. }k^{\left. \beta \right\rangle }\sum_{n=0}^{\infty
}a_{n}(\omega) E_{\mathbf{k}}^{n}.  \label{expansion}
\end{equation}%
Substituting Eq.\ (\ref{expansion}) into Eq.\ (\ref{Exp}), we obtain the
following expression for $\tilde{G}_{R}^{\mu \nu \alpha \beta }( \omega
,\mathbf{0}) $, 
\begin{eqnarray}
\tilde{G}_{R}^{\mu \nu \alpha \beta }( \omega ,\mathbf{0})
&=&\sum_{n=0}^{\infty }a_{n}( \omega) \int dK\,k^{\left\langle \mu
\right. }k^{\left. \nu \right\rangle }k^{\left\langle \alpha \right.
}k^{\left. \beta \right\rangle }E_{\mathbf{k}}^{n}f_{0\mathbf{k}} \notag \\ 
&=&2\Delta^{\mu \nu \alpha \beta }
\sum_{n=0}^{\infty }I_{n+4,2}\,a_{n}( \omega)\;,
\end{eqnarray}%
where we introduced the thermodynamic integral%
\begin{equation}
I_{nq}=\frac{1}{\left( 2q+1\right) !!}\int dK\,f_{0\mathbf{k}}E_{\mathbf{k}%
}^{n-2q}\left( m^{2}-E_{\mathbf{k}}^{2}\right) ^{q}.
\end{equation}

Thus, the relation between $\tilde{\pi}^{\mu \nu }$ and $\tilde{\sigma}^{\mu
\nu }$ can be cast into a more convenient form $\tilde{\pi}^{\mu \nu }\left(
\omega ,\mathbf{0}\right) =2\tilde{G}_{R}\left( \omega ,\mathbf{0}\right) 
\tilde{\sigma}^{\mu \nu }\left( \omega ,\mathbf{0}\right) $, with $\tilde{G}%
_{R}\left( \omega ,\mathbf{0}\right) $ being given by
\begin{equation}
\tilde{G}_{R}( \omega ,\mathbf{0}) =\sum_{n=0}^{\infty
}\,I_{n+4,2}\,a_{n}( \omega ) \text{.}
\end{equation}

The whole frequency dependence of $\tilde{G}_{R}( \omega ,\mathbf{0}) $
is contained in the functions $a_{n}( \omega ) $. The problem is
then reduced to determining the analytic properties of $a_{n}( \omega) $. 
The way to solve this problem is to substitute the expansion (\ref%
{expansion}) into Eq.\ (\ref{equation}), multiply by $E_{\mathbf{k}%
}^{m}k^{\left\langle \mu \right. }k^{\left. \nu \right\rangle }$, and
integrate over $dK$. Then one obtains%
\begin{equation}
\sum_{n=0}^{\infty }\left( -i\omega \mathcal{D}^{mn}+\mathcal{A}^{mn}\right)
a_{n}\left( \omega \right) =\beta _{0}I_{m+3,2}\;,
\end{equation}%
where we defined the matrices 
\begin{eqnarray}
\mathcal{A}^{mn}\Delta ^{\mu \nu \alpha \beta } &=&-\frac{1}{2}\int dKE_{%
\mathbf{k}}^{m}k^{\left\langle \mu \right. }k^{\left. \nu \right\rangle }%
\hat{C}\left( K\right) f_{0\mathbf{k}}E_{\mathbf{k}}^{n}k^{\left\langle
\alpha \right. }k^{\left. \beta \right\rangle }, \\
\mathcal{D}^{mn} &=&\frac{1}{5!!}\int dK\,f_{0\mathbf{k}}E_{\mathbf{k}%
}^{m+n}\left( m^{2}-E_{\mathbf{k}}^{2}\right) ^{2}\,.
\end{eqnarray}%
The solution of this equation can be formally expressed as, 
\begin{equation}
a_{m}( \omega ) =\beta _{0}\sum_{n=0}^{\infty }\left[ \left(
-i\omega \mathcal{D}+\mathcal{A}\right) ^{-1}\right] ^{mn}I_{n+3,2}.
\end{equation}%
Thus, we obtain 
\begin{equation}
\tilde{G}_{R}( \omega ,\mathbf{0}) =\beta _{0}\sum_{m=0}^{\infty
}\sum_{n=0}^{\infty }I_{m+4,2}\left[ \left( -i\omega \mathcal{D}+\mathcal{A}%
\right) ^{-1}\right] ^{mn}I_{n+3,2}. 
\end{equation}
The poles of this function can be obtained from the roots of the
determinant $\det \left( -i\omega \mathcal{D}+\mathcal{A}\right) =0$. Thus,
the relaxation times and viscosity coefficients are determined by inverting
and finding eigenvalues of infinite-dimensional matrices. In practice,
however, the expansion (\ref{expansion}) has to be truncated and one never
actually deals with infinite matrices. This is a common approach and happens
quite often when dealing with the collision operator. Let us try the
simplest case and consider only one term in the expansion (this corresponds
to using the Israel-Stewart 14-moment approximation in the moments method).
Then all matrices become numbers and everything simplifies considerably.
In this case, the retarded Green's function has only one pole, $\omega
_{0}=-i\mathcal{A}^{00}/I_{42}$ \cite{DNNR}.

Since in this case the thermodynamic force is given by $F=2 
\sigma^{\mu \nu }$, the shear viscosity coefficient is given by $D_{0}$. 
Thus, the shear viscosity and relaxation time are given by%
\begin{eqnarray}
\eta  &=&\tilde{G}_{R}( 0 ,\mathbf{0}) =\beta _{0}\frac{%
I_{42}I_{32}}{\mathcal{A}^{00}}, \\
\tau _{\pi } &=&\frac{1}{i\omega _{0}}=\frac{I_{42}}{\mathcal{A}^{00}}.
\end{eqnarray}%
This implies that $\eta /\tau _{\pi }=\beta _{0}I_{32}$. In the massless limit,
for a gas of hard spheres, $%
\mathcal{A}^{00}$ $=\left( 3/5\right) I_{42}\,n_{0}\,\sigma $, with $\sigma $
being the total cross-section.

Exactly the same results were obtained in Ref.\ \cite{dkr} within a completely
different approach. This clearly demonstrates that the relaxation time in
Israel-Stewart theories, which determines the time scale related to
transient dynamics, is indeed a microscopic scale. In other words, the
relaxation time is determined by the inter-particle scattering rate and not
by an arbitrary fluid-dynamical time scale.

\section{Conclusion}

In this work we proved that the existence of singularities in the retarded
Green's functions considerably restricts the applicability of the gradient
expansion as a method to extract macroscopic dynamics. As a matter of fact,
the causal structure of relativistic fluid dynamics hinges on the
presence of poles in the Green's functions.

We computed the first pole of the retarded Green's function associated with
shear stress in a weakly coupled classical gas of hard spheres. We proved
that the inverse of this pole is directly related to the relaxation time
that appears in Israel-Stewart theory. Consequently, the relaxation time in
Israel-Stewart theories corresponds to a microscopic time scale and such
theories are indeed describing the real transient dynamics of the fluid.

In summary, the true relaxation time is always given by the first pole of
the retarded Green's function. In general, the location of this pole cannot
be found using a truncated Taylor expansion around the origin or, in other
words, via the gradient expansion. This result should be expected since the
transient behavior of a given system cannot be unambiguously extracted from
its asymptotic solution given by the gradient expansion.

The authors thank T.~Kodama, T.~Koide, A.~Ficnar, and G.~Torrieri for
insightful discussions. H.N.\ was supported by the ExtreMe Matter Institute
EMMI. The authors thank the Helmholtz International Center for FAIR within
the framework of the LOEWE program for support.

\end{document}